\documentstyle[11pt]{article}\textheight 230mm\textwidth 150mm
            \newcommand{\be}{\begin{eqnarray}}
            \newcommand{\ee}{\end{eqnarray}}
            \newcommand{\eel}[1]{\label{#1}\end{eqnarray}}
\newcommand{\e}[1]{\label{e:#1}\end{eqnarray}}
     \newcommand{\eg}{{\em e.g.\ }}
            \newcommand{\ie}{{\em i.e.\ }}

            \newcommand{\la}{{\lambda}}
            
            \newcommand{\del}{{\delta}}

\newcommand{\cM}{{\cal{M}}}

\newcommand{\cW}{{\cal{W}}}
\newcommand{\cX}{{\cal{X}}}
           \newcommand{\ra}{{\rightarrow}}
 \newcommand{\lea}{{\leftarrow}}
            
            \newcommand{\Lra}{{\Leftrightarrow}}
            \newcommand{\pet}{{\cal P}}

\newcommand{\ca}{{\cal C}}

            \newcommand{\beq}{\begin{quote}}
            \newcommand{\eq}{\end{quote}}
            \newcommand{\Om}{\Omega}
   
            \newcommand{\al}{\alpha}
            \newcommand{\ben}{\begin{enumerate}}
            \newcommand{\een}{\end{enumerate}}
            \newcommand{\bit}{\begin{itemize}}
            \newcommand{\ei}{\end{itemize}}
    	\newcommand{\nn}{\nonumber}
            \newcommand{\r}[1]{(\ref{e:#1})}
            \newcommand{\edfl}[1]{\label{#1}\end{df}}
\def\theequation{\thesection.\arabic{equation}}
\newcommand{\vb}{{\cal h}}
\newcommand{\hb}{{\cal i}}

\newcommand{\Ra}{{\Rightarrow}}
\newcommand{\ve}{{\varepsilon}}

\newcommand{\dagg}{^{\dag}}

\newcommand{\bett}{{\bf 1}}
	
\def\d{\partial}
\def\cC{{\cal C}}

 \def\cH{{\cal H}}

  \def\half{{1 \over 2}}

\begin{document}
\begin{titlepage}
\noindent
G\"{o}teborg ITP 99-04\\
May 12, 1999\\

\vspace*{5 mm}
\vspace*{35mm}
\begin{center}{\LARGE\bf General quantum antibrackets}
\end{center} \vspace*{3 mm} \begin{center} \vspace*{3 mm}

\begin{center}Igor Batalin\footnote{On leave of absence from
P.N.Lebedev Physical Institute, 117924  Moscow, Russia\\E-mail:
batalin@td.lpi.ac.ru.} and Robert
Marnelius\footnote{E-mail: tferm@fy.chalmers.se.}\\
\vspace*{7 mm} {\sl Institute
of Theoretical Physics\\ Chalmers University of
Technology\\ G\"{o}teborg
University\\ S-412 96  G\"{o}teborg, Sweden}\end{center}
\vspace*{25 mm}
\begin{abstract}
The recently introduced quantum antibracket is further generalized allowing
for the
defining odd operator Q to be arbitrary. We give exact formulas for higher
quantum antibrackets of arbitrary orders and their generalized Jacobi
identities.
Their applications to BV-quantization and BFV-BRST quantization are then
reviewed
including some new aspects.
\end{abstract}\end{center}\end{titlepage}

\setcounter{page}{1}
\section{Introduction}
In \cite{Quanti} we
introduced new quantum objects called quantum antibrackets, which are  operator
mappings from  classical antibrackets exactly like commutators are mappings from
Poisson brackets. Classical antibrackets were introduced in
\cite{anti,BV} and have mainly been used in the Lagrangian BV-quantization
 of gauge theories  \cite{BV}. Apart from
providing an operator version of the BV-quantization \cite{Quanti}, the quantum
antibrackets have been used to give a new kind of quantum master equation for
generalized quantum Maurer-Cartan equations for arbitrary open groups \cite{OG}.
Remarkably enough the quantum antibrackets used in these master equations were
generalized ones, which neither satisfy  the Jacobi identities nor
Leibniz' rule.
For such brackets one has to use a hierarchy of higher quantum antibrackets
defined
in a definite way. In their first form they were introduced already in
\cite{Quanti}
which is valid for the restricted case when all operators commute.  A more
general
form valid for  operators in arbitrary involutions were  given in
\cite{OG}. (In
\cite{Dual} it was also shown that all these results have, at the classical
level, a
dual version in terms of new types of generalized Poisson brackets.) In
\cite{Sp2QA}
the quantum antibrackets were generalized to Sp(2)-brackets defined
 in analogy to the Sp(2)-antibrackets used in the Sp(2)-version of
BV-quantization
\cite{Sp2, Sp2s,Trip} and most of the above results were then generalized.

Here we review these results with some natural further developments. We
start from the most general quantum antibracket as defined in \cite{Quanti}  and
give then systematically  all their main properties. Finally we review the
applications considered so far.  The quantum antibrackets and their exact
properties are given in sections 2 and 3, and the main part of the presented
results are new. In section 4 we consider then the restriction to ordinary
quantum
antibrackets which satisfy all properties as required by the conventional
classical
antibrackets. In sections 5 and 6 we review the applications to
BV-quantization and
BFV-BRST quantization respectively including some new aspects. Finally, in
section 7
we give some remarks on the corresponding properties and applications of quantum
Sp(2)-antibrackets.

\section{General quantum antibrackets}
The general quantum antibracket is defined by the expression \cite{Quanti}
\be
&&(f, g)_Q\equiv\half \left([f, [Q, g]]-[g, [Q,
f]](-1)^{(\ve_f+1)(\ve_g+1)}\right),
\e{1}
where  $f$ and $g$ are any operators with Grassmann parities
$\ve(f)\equiv\ve_f$ and
$\ve(g)\equiv\ve_g$ respectively. $Q$ is an odd operator, $\ve(Q)=1$.  The
commutators
on the right-hand side is the graded commutator defined by
\be
&&[f, g]\equiv fg-gf(-1)^{\ve_f\ve_g}, \quad \forall f,g.
\e{2}
 The quantum antibracket \r{1} satisfies the
properties:\\
\\ 1) Grassmann parity
\be
&&\ve((f, g)_Q)=\ve_f+\ve_g+1.
\e{3}
2) Symmetry
\be
&&(f, g)_Q=-(g, f)_Q(-1)^{(\ve_f+1)(\ve_g+1)}.
\e{4}
3) Linearity
\be
&&(f+g, h)_Q=(f, h)_Q+(g, h)_Q, \quad (\mbox{for}\; \ve_f=\ve_g).
\e{5}
4) If one entry is an odd/even parameter $\la$ we have
\be
&&(f, \la)_Q=0\quad {\rm for\ any\ operator}\;f.
\e{6}
All these properties agree exactly with the corresponding properties of the
classical
antibracket $(f,g)$ for functions $f$ and $g$. However, the classical
antibracket
satisfies in addition \\ \\
 5) the Jacobi identities
\be
&&(f,(g, h))(-1)^{(\ve_f+1)(\ve_h+1)}+cycle(f,g,h)\equiv0,
\e{7}
6) and  Leibniz' rule
\be
&&(fg, h)-f(g, h)-(f, h)g(-1)^{\ve_g(\ve_h+1)}=0.
\e{8}
These properties are not satisfied by the quantum antibracket \r{1} in its
general
form. Instead we have\\
\be
&&(f,(g, h)_Q)_Q(-1)^{(\ve_f+1)(\ve_h+1)}+cycle(f,g,h)=\nn\\&&={1\over
6}(-1)^{\ve_f+\ve_g+\ve_h}\left\{\left([f, [g, [h, Q^2]]]+\half[[f, [g,
[h, Q]]], Q]\right)(-1)^{\ve_f\ve_h}+\right.\nn\\
&&+\left.\left([f, [h,
[g, Q^2]]]+\half[[f, [h,
[g, Q]]],
Q]\right)(-1)^{\ve_h(\ve_f+\ve_g)}\right\}+cycle(f,g,h),
\e{9}
and
\be
&&(fg, h)_Q-f(g, h)_Q-(f, h)_Qg(-1)^{\ve_g(\ve_h+1)}=\nn\\
&&=\half\left([f, h][g,
Q](-1)^{\ve_h(\ve_g+1)}+[f,Q][g,h](-1)^{\ve_g}\right).
\e{10}

In our previous treatments we have only considered nilpotent $Q$-operators
which is a
natural restriction from the point of view of BV-quantization. Here we
leave $Q$ unrestricted in the general treatment.  One may then notice that
any odd
operator
$Q$ satisfies the algebra
\be
&&[Q, Q]=2Q^2,\quad [Q^2, Q^2]=[Q^2, Q]=0,
\e{11}
which directly follows from the definition \r{2}.
In terms of the quantum antibracket \r{1} $Q$ satisfies the
algebra
\be
&&(Q, Q)_Q=(Q^2, Q^2)_Q=(Q^2, Q)_Q=0.
\e{12}
Notice also the relation
\be
&&(f, Q)_Q={3\over 2}[f, Q^2], \quad \forall f.
\e{13}
A nonzero $Q^2$ complicates some formulas. For instance,  instead of
(20) in \cite{Quanti} we have here
\be
&&[Q, (f, g)_Q]=([Q, f], g)_Q+(f, [Q,
g])_Q(-1)^{\ve_f+1}-\nn\\&&-(-1)^{\ve_f}\left([f, [g, Q^2]]+[g, [f,
Q^2]](-1)^{\ve_f\ve_g}\right)=[[Q,f],
[Q, g]]-\nn\\ &&-\half(-1)^{\ve_f}\left([f, [g, Q^2]]+[g, [f,
Q^2]](-1)^{\ve_f\ve_g}\right).
\e{14}
In our derivation of the generalized Jacobi identities \r{9} we have used
the relation
\be
&&(f, (g, h)_Q)_Q=[[f, Q], (g, h)_Q]]+\half[Q, [f, (g, h)_Q]](-1)^{\ve_f},
\e{15}
which directly follows from the definition \r{1} (further details are given
below).

\setcounter{equation}{0}
\section{Hierarchy of general  higher order
antibrackets and\\ generalized Jacobi identities.}
Although the quantum antibracket \r{1} in its general form does not satisfy the
Jacobi identities and Leibniz' rule, we may provide for a systematic
description of
its algebraic properties in terms of higher order quantum antibrackets.

In \cite{Quanti,OG} it was shown that higher order quantum antibrackets in terms
of operators in involution may be defined in terms of a generating operator
and that
the generating operator determine the Jacobi identities. This construction
may in
fact be generalized to such an extent that it allows us to define general
higher order
quantum antibrackets in terms of {\em arbitrary} operators. For the general
case  we
need then the following two generating operators (In refs.\cite{Quanti,OG} $Q$
was considered to be nilpotent, $Q^2=0$.)
\be
&&Q(\la)\equiv e^{-A}Qe^A, \quad Q^2(\la)\equiv e^{-A}Q^2e^A,
\e{201}
where $A$ is an even operator defined by
\be
&&A=f_a\la^a,
\e{202}
where  $f_a$, $a=1,2,\ldots$, are arbitrary operators with Grassmann
parities $\ve_a\equiv \ve(f_a)$ and where $\la^a$ are parameters with Grassmann
parities $\ve(\la^a)=\ve_a$. (In \cite{Quanti} $\{f_a\}$ was a set of commuting
operators.) We have the equalities
\be
&&Q(\la)=\sum_{n=0}^\infty Q_n, \quad Q^2(\la)=\sum_{n=0}^\infty Q^2_n,
\e{203}
where
\be
&&Q_0\equiv Q,\quad Q_n\equiv {1\over n!}[\cdots[[Q, A],\cdots,A]_n \; {\rm for\
}n\geq1,\nn\\
&&Q^2_0\equiv Q^2,\quad Q^2_n\equiv {1\over n!}[\cdots[[Q^2, A],\cdots,A]_n
\; {\rm
for\ }n\geq1,
\e{204}
where the last index $n$ indicates that the expression involves $n$
commutators.
We define the general higher order quantum antibrackets by (cf \cite{Quanti})
\be
&&(f_{a_1},\ldots, f_{a_n})_Q\equiv -
\left.Q(\la)\stackrel{\lea}{\d}_{a_1}\stackrel{\lea}{\d}_{a_2}\cdots
\stackrel{\lea}{\d}_{a_n}(-1)^{E_n}\right|_{\la=0}=\nn\\&&=-
Q_n\stackrel{\lea}{\d}_{a_1}\stackrel{\lea}{\d}_{a_2}\cdots
\stackrel{\lea}{\d}_{a_n}(-1)^{E_n}=\nn\\&&=
-{1\over n!}[\cdots[[Q,
f_{b_1}],\cdots,f_{b_n}]\la^{b_n}\cdots\la^{b_1}
\stackrel{\lea}{\d}_{a_1}\stackrel{\lea}{\d}_{a_2}\cdots
\stackrel{\lea}{\d}_{a_n}(-1)^{E_n},\nn\\&&
\quad E_n\equiv
\sum_{k=0}^{\left[{n-1\over 2}\right]}\ve_{a_{\rm 2k+1}},
\e{205}
where $\stackrel{\lea}{\d}_{a}\equiv \stackrel{\lea}{\d}/\d\la^a$. This
definition
yields higher order quantum antibrackets for arbitrary operators in terms of
symmetrized multiple commutators with an arbitrary odd $Q$:
\be
&&(f_{a_1},\ldots, f_{a_n})_Q= -{1\over
n!}(-1)^{E_n}\sum_{\rm sym} [\cdots[[Q, f_{a_1}],\cdots,f_{a_n}],
\e{206}
where the sum is over all possible orders of $a_1, \ldots, a_n$ ($n!$
terms) with
appropriate sign factors obtained from the corresponding reordering of the
monomial
$\la^{a_1}\cdots\la^{a_n}$ (this follows from the last equality in \r{205}).
The higher antibrackets satisfy  the properties
\be
&&(\ldots, f_a, f_b, \ldots)_{Q}=
-(-1)^{(\ve_a+1)(\ve_b+1)}
(\ldots, f_b, f_a, \ldots)_{Q},\nn\\
&&\ve((f_{a_1},
f_{a_2},\ldots,f_{a_n})_{Q})=\ve_{a_1}+
\cdots+\ve_{a_n}+1.
\e{2061}
To the lowest orders  we get explicitly
\be
&&n=2:\quad (f_a, f_b)_Q=\half \left([f_a, [Q, f_b]]-[f_b, [Q,
f_a]](-1)^{(\ve_a+1)(\ve_b+1)}\right),\nn\\
&&n=3: \quad (f_a, f_b,
f_c)_Q={1\over6}(-1)^{(\ve_a+1)(\ve_c+1)}\left([[[Q,f_a],f_b],
f_c](-1)^{\ve_a\ve_c}+\right.\nn\\&&\left.+[[[Q, f_c], f_b],
f_a](-1)^{\ve_b(\ve_a+\ve_c)}+cycle(a,b,c)\right),
\e{207}
where the expression for $n=2$ is exactly the antibracket \r{1}. For
nilpotent $Q$
and commuting operators $f_{a_k}$ \r{206} reduces to the higher quantum
antibrackets
in \cite{Quanti} (eq.(33)). If furthermore $f_{a_k}$ are functions of some
coordinates
and
$Q$ a nilpotent differential operator then \r{206} on unity
yields the classical higher antibrackets considered in \cite{hanti}.

The higher order quantum antibrackets may also be expressed recursively in
terms of
the next lower ones. The precise relations may be obtained from the
recursion relation
\be
&&Q_n={1\over n}[Q_{n-1}, A].
\e{208}
This inserted into \r{205} yields
\be
&&(f_{a_1},\ldots, f_{a_n})_Q={1\over n}\sum_{k=1}^n
[(f_{a_1},\ldots,f_{a_{k-1}},
f_{a_{k+1}}, \ldots, f_{a_{n}})_Q, f_{a_k}](-1)^{B_{k,n}}=\nn\\&&=-{1\over
n}\sum_{k=1}^n [ f_{a_k}, (f_{a_1},\ldots,f_{a_{k-1}}, f_{a_{k+1}}, \ldots,
f_{a_{n}})_Q](-1)^{C_{k,n}},
\e{209}
where
\be
&&B_{k,n}\equiv\ve_{a_k}(\ve_{a_{k+1}}+\ldots
+\ve_{a_n})+\sum_{s=2[{k\over2}]+1}^n\ve_{a_s},\nn\\
&&C_{k,n}\equiv\ve_{a_k}(\ve_{a_1}+\ldots
+\ve_{a_{k-1}})+\ve_{a_k}+\sum_{s=2[{k\over2}]+1}^n\ve_{a_s}.
\e{2091}
 To the lowest
order we have explicitly (this expression was also given in \cite{OG})
\be
 &&(f_a, f_b, f_c)_Q={1\over
3}(-1)^{(\ve_a+1)(\ve_c+1)}\left([(f_a, f_b)_Q,
f_c](-1)^{\ve_c+(\ve_a+1)(\ve_c+1)}+cycle(a,b,c)\right)=\nn\\&&={1\over
3}(-1)^{(\ve_a+1)(\ve_c+1)}\left([f_a, (f_b,f_c)_Q]
(-1)^{\ve_b+\ve_a(\ve_c+1)}+cycle(a,b,c)\right).
\e{210}

In order to derive generalized Jacobi identities we notice first that the
definition
\r{1} of the antibracket yields the relation (cf \r{15})
\be
&&(f_{a_k}, (f_{a_1},\ldots,f_{a_{k-1}}, f_{a_{k+1}}, \ldots, f_{a_n})_Q)_Q=
[[f_{a_k}, Q], (f_{a_1},\ldots,f_{a_{k-1}}, f_{a_{k+1}}, \ldots,
f_{a_n})_Q]+\nn\\&&+\half[Q, [f_{a_k}, (f_{a_1},\ldots,f_{a_{k-1}}, f_{a_{k+1}},
\ldots, f_{a_n})_Q](-1)^{\ve_{a_k}}.
\e{2101}
This combined with the recursion relation
 \r{209} implies then
\be
&&\sum_{k=1}^n(f_{a_k}, (f_{a_1},\ldots,f_{a_{k-1}}, f_{a_{k+1}}, \ldots,
f_{a_n})_Q)_Q(-1)^{D_{k,n}}=\nn\\&&=\sum_{k=1}^n[[f_k, Q],
(f_{a_1},\ldots,f_{a_{k-1}}, f_{a_{k+1}}, \ldots,
f_{a_n})_Q](-1)^{D_{k,n}}-\nn\\&&-{n\over2}[Q, (f_{a_1},\ldots,
f_{a_n})_Q], \nn\\&&
D_{k,n}=\ve_{a_k}+C_{k,n}= \ve_{a_k}(\ve_{a_1}+\ldots
+\ve_{a_{k-1}})+\sum_{s=2[{k\over2}]+1}^n\ve_{a_s}.
\e{2102}
The first sum on the right-hand side may be expressed in terms of higher
antibrackets
 by means of the
 identities  (cf
\cite{Quanti})
\be
&&\left.\left([Q(\la), Q(\la)]-2Q^2(\la)\right)\stackrel{\lea}{\d}_{a_1}
\stackrel{\lea}{\d}_{a_2}\cdots
\stackrel{\lea}{\d}_{a_n}\right|_{\la=0}=0,
\e{211}
which are equivalent to
\be
&&
\left(\sum_{k=0}^n[Q_k,
Q_{n-k}]-2Q^2_n\right)\stackrel{\lea}{\d}_{a_1}\stackrel{\lea}{\d}_{a_2}\cdots
\stackrel{\lea}{\d}_{a_n}=0,
\e{212}
where $Q^2_n$ is defined in \r{204}.
For $n=0,1,2,3$ we have explicitly
\be
&&n=0:\quad [Q,Q]-2Q^2=0,\nn\\
&&n=1:\quad [Q, [Q, f_a]]-[Q^2, f_a]=0,\nn\\
&&n=2:\quad [Q, (f_a, f_b)_Q]-[[Q, f_a], [Q,
f_b]]+\nn\\&&\quad\quad+\half[[Q^2, f_a],
f_b](-1)^{\ve_a}+\half[[Q^2, f_b], f_a](-1)^{\ve_b(\ve_a+1)}=0,\nn\\
&&n=3:\quad [Q, (f_a, f_b,
f_c)_Q](-1)^{(\ve_a+1)(\ve_c+1)}+\nn\\&&\quad\quad+\left([[f_a, Q], (f_b,
f_c)_Q](-1)^{\ve_b+\ve_a\ve_c}+cycle(a,b,c)\right)-\nn\\&&\quad\quad-{1\over
6}\sum_{\rm
sym}[[[Q^2, f_a], f_b], f_c](-1)^{\ve_a\ve_c}=0.
\e{213}
The first is just a trivial identity, the second is a Jacobi identity, and
the third
is exactly \r{14}.  For $n\geq3$  \r{212} multiplied by $(-1)^{E_n}$ becomes
\be
&&-[Q, (f_{a_1},\ldots,
f_{a_n})_Q]+\sum_{k=1}^n[[f_{a_k}, Q], (f_{a_1},\ldots,f_{a_{k-1}},
f_{a_{k+1}}, \ldots, f_{a_n})_Q](-1)^{D_{k,n}}+\nn\\&&+R_n-(-1)^{E_n}{1\over
n!}\sum_{\rm sym}[\cdots[[Q^2,f_{a_1}], f_{a_2}],\cdots, f_{a_n}]=0,
\e{2131}
where
\be
&&R_n\equiv \half\sum_{k=2}^{n-2}[Q_k,
Q_{n-k}]\stackrel{\lea}{\d}_{a_1}\stackrel{\lea}{\d}_{a_2}\cdots
\stackrel{\lea}{\d}_{a_n}(-1)^{E_n}=\nn\\&&=
\half\sum_{k=2}^{n-2}\sum_{\rm sym}[(f_{a_1},\ldots, f_{a_k})_Q,
(f_{a_{k+1}}, \ldots,
f_{a_n})_Q](-1)^{F_{k,n}}, \quad F_{k,n}\equiv \sum_{r=1}^{(n,k)}\ve_{a_r},
\e{21311}
where $(n,k)\equiv n$ for $k$ odd, and $(n,k)\equiv k$ for $k$ even. The
symmetrized
sum is over all different orders with additional sign factors
$(-)^{E_n+\tilde{E}_n+A_n}$ where $\tilde{E}_n$ is $E_n$ for the new order
and $A_n$
from the reordering of the monomial $\la^{a_1}\cdots\la^{a_n}$.

 The expression \r{2131} inserted into \r{2102} leads then
to the generalized Jacobi identities
\be
&&\sum_{k=1}^n(f_{a_k}, (f_{a_1},\ldots,f_{a_{k-1}}, f_{a_{k+1}}, \ldots,
f_{a_n})_Q)_Q(-1)^{D_{k,n}}=-{n-2\over2}[Q, (f_{a_1},\ldots,
f_{a_n})_Q]-\nn\\&&-R_n+(-1)^{E_n}{1\over n!}\sum_{\rm
sym}[\cdots[[Q^2,f_{a_1}], f_{a_2}],\cdots, f_{a_n}]=0
\e{2132}
where
 $R_n$ is the sum of commutators of antibrackets of orders between $2$ and
$n-2$ given
by \r{21311}.  For
$n=3$ we have in particular  ($R_3=0$)
\be
&&(f_a, (f_b, f_c)_Q)_Q(-1)^{(\ve_a+1)(\ve_c+1)}+cycle(a,b,c)=
-\half[(f_a, f_b,
f_c)_Q(-1)^{(\ve_a+1)(\ve_c+1)},
Q]-\nn\\&&-{1\over6}(-1)^{\ve_a+\ve_b+\ve_c}\left(
[[[Q^2, f_a], f_b], f_c](-1)^{\ve_a\ve_c}+[[[Q^2, f_c], f_b],
f_a](-1)^{\ve_b(\ve_a+\ve_c)}\right.+\nn\\&&+\left.cycle(a,b,c)\right),
\e{215}
which may be rewritten as \r{9}.

\setcounter{equation}{0}
\section{Ordinary quantum antibrackets}
It is natural to impose a restriction such that  the
 quantum antibrackets  satisfy the Jacobi identities
\r{7} and  Leibniz' rule \r{8}. Such antibrackets we  call ordinary quantum
antibrackets.   This requires of course that all higher order quantum
antibrackets
vanish (from $n=3$). The precise conditions are best extracted from
the explicit expressions \r{9} and \r{10}.    From
\r{10} we find that the quantum antibracket
\r{1} satisfies  Leibniz' rule \r{8}, \ie
\be
&&(fg, h)_Q-f(g, h)_Q-(f, h)_Qg(-1)^{\ve_g(\ve_h+1)}=0,
\e{101}
for two classes of operators: the class of commuting operators or the class of
operators commuting with $Q$. Since the antibracket is zero in the latter
case we
consider  only  maximal sets of commuting operators denoted $\cM$ from now
on.

The Jacobi identities \r{7} are  satisfied provided the operator $Q$ satisfies
the condition
\be
&&[f, [g, [h, Q^2]]]+\half[[f, [g,
[h, Q]]], Q]=0, \quad \forall f,g,h\in\cM.
\e{1011}
(This condition is trivially satisfied  for the class of operators
commuting with
$Q$.)  First we notice that for the class of commuting
operators,
$\cM$, the quantum antibracket
\r{1} reduces to
\be
&&(f, g)_Q=[f, [Q, g]]=[[f, Q], g], \quad \forall f,g\in\cM.
\e{105}
In order to use this expression repeatedly like in the Jacobi
identities, we must require
\be
&&(f, g)_Q\in\cM, \quad \forall f,g\in\cM,
\e{104}
which is equivalent to
\be
&&[f, [g, [h, Q]]]=0,  \quad \forall f,g,h\in\cM.
\e{102}
However, this together with \r{1011} requires then
\be
&&[f, [g, [h, Q^2]]]=0, \quad \forall f,g,h\in\cM.
\e{103}
Thus, ordinary nontrivial quantum antibrackets are defined for a class of
commuting
operators $\cM$. It is naturally defined by \r{105} where $Q$   must satisfy the
conditions
\r{102} and \r{103}.

In order to give explicit solutions we consider like in \cite{Quanti,Dual} a
supersymmetric manifold of dimension $(2n,2n)$ spanned by the canonical
coordinates
$\{x^a, x^*_a, p_a, p^a_*\}$, $a=1,\ldots,n$,   with Grassmann parities
$\ve_a\equiv\ve(x^a)=\ve(p_a)$ and $\ve(x^*_a)=\ve(p^*_a)=\ve_a+1$.
Their canonical commutation relations have the nonzero part
\be
&&[x^a, p_b]=i\hbar\del^a_b,\quad [x^*_a, p^b_*]=i\hbar\del^b_a.
\e{106}
Let now the class of commuting operators $\cM$ be all functions of $x^a$
and $x^*_a$.
The following $Q$-operator satisfies then the conditions \r{102} and \r{103}:
\be
&&Q=p_a p^a_*(-1)^{\ve_a}.
\e{107}
In terms of this $Q$ the quantum antibracket \r{1} becomes
\be
&&(i\hbar)^{-2}(f, g)_Q=f\stackrel{\lea}{\d}_a
\d_*^ag-g\stackrel{\lea}{\d}_a\d_*^a f(-1)^{(\ve_f+1)(\ve_g+1)},
\e{108}
which exactly agrees with the standard classical antibracket in which $x^a$ and
$x^*_a$ are fields and antifields \cite{BV}. The $Q$-operator \r{107} is
nilpotent
($Q^2=0$).
However, the $Q$-operator is not uniquely determined by the quantum
antibracket. In
fact, the quantum antibracket \r{108} is also obtained from the $Q$-operator
\be
&&Q=p_a p^a_*(-1)^{\ve_a}+i\hbar
f^a(x,x^*)p_a(-1)^{\ve_a}+i\hbar
f^*_a(x,x^*)p^a_*(-1)^{\ve_a+1}-(i\hbar)^2g(x,x^*)\nn\\
\e{112}
for any functions $f^a$, $f^*_a$ and $g$ of the commuting operators $x^a$ and
$x^*_a$. (The normalization chosen in \r{112} is appropriate for applications to
BV-quantization.) The
$Q$-operator \r{112} is in general not nilpotent.

In terms of general coordinates $X^A$,
$A=1,\ldots,2n$, the solution of the condition \r{102} is
\be
&&Q=-\half  E^{AB}(X)P_BP_A(-1)^{\ve_B}+i\hbar
F^A(X)P_A(-1)^{\ve_A}-(i\hbar)^2G(X),
\e{113}
where $P_A$ ($\ve(P_A)=\ve(X^A)\equiv\ve_A$) is the conjugate momentum
operator to
$X^A$ ($[X^A, P_B]=i\hbar\del_B^A$). In order to also satisfy condition \r{103},
$Q^2$ must be at most quadratic in the conjugate momenta. This requires
\be
&&E^{AD}\d_DE^{BC}(-1)^{(\ve_A+1)(\ve_C+1)}+cycle(A,B,C)=0.
\e{114}
The $Q$-operator \r{113} inserted into \r{1} for the class of operators
which are
functions of $X^A$ yields the quantum antibracket
\be
&&(i\hbar)^{-2}(f, g)_Q=f\stackrel{\lea}{\d}_A E^{AB}\d_{B}\, g, \quad \forall
f,g\in\cM,
\e{115}
where $\d_A\equiv
\d/\d X^A$. One may easily check that the Jacobi identities of \r{115} requires
\r{114} (cf
\cite{BTU}). The operator \r{113} is like \r{112} in general not nilpotent.
(In fact,
\r{112} is \r{113} in Darboux coordinates $x^a$ and $x_a^*$.) Notice that if $Q$
has terms involving
third or higher powers of
$P_A$ then even the general condition \r{1011} does not allow for any
solutions since
the first term in
\r{1011} involves first order derivatives of $f, g, h$ while the second
term involves
second derivatives.

\setcounter{equation}{0}
\section{Operator version of BV-quantization}
The Lagrangian BV-quantization of general gauge theories is formulated
within the
path integral formulation. The classical antibrackets play a crucial role
in this
formulation. Here we show that the quantum antibrackets provide for the
appropriate
algebraic tools in a corresponding operator formulation. (These results
were given in
\cite{Quanti,Sp2QA}.) In this construction it is the ordinary quantum
antibrackets
that provide for the relevant framework. We have therefore to restrict
ourselves to a
maximal commuting set of operators. This is just a polarization into fields
and their
momenta. The field operators are then the maximal set to be chosen. The crucial
ingredient in the BV-quantization is the quantum master equation
\be
&&\Delta e^{{i\over\hbar}\cW}=0,
\e{301}
where $\Delta$ is an odd, second order differential operator. $\cW$ is the
master
action. Within the operator formulation \r{301} is replaced by
\be
&&Q|\cW\hb=0,
\e{302}
where $Q$ is the odd operator that enters the definition of the quantum
antibracket.
The BV-formulation requires $Q$ to be
\ben
\item maximally quadratic in the momenta
\item nilpotent ($Q^2=0$)
\item hermitian
\item reduce to the differential operator $\Delta$ in the Schr\"odinger
picture which
should have no $\hbar$-dependence.
\een
The first and last condition requires $Q$ to be of the form \r{112} or
\r{113} with
$\hbar$-independent functions.  Starting from the general $Q$-operator
\r{113}  we
find then the general hermitian, nilpotent solution
\be
&&Q\equiv-\half\rho^{-1/2}P_A\rho
E^{AB}P_B\rho^{-1/2}(-1)^{\ve_B}-(i\hbar)^2G(X),
\e{303}
where $\rho(X)$ is the volume density operator. $Q$ is hermitian if $\rho,
X^A, G$
and $E^{AB}$ are hermitian, and $P^{\dag}_A=(-1)^{\ve_A}P_A$. The
$\hbar$-dependence
in $Q$ is chosen such that
\be
&&\vb X|Q|\cW\hb=-(i\hbar)^2\Delta\vb X|\cW\hb,
\e{304}
where
\be
&&\Delta=\Delta_0+G(X),  \quad \Delta_0\equiv\half\rho^{-1}\d_A\circ\rho
E^{AB}\d_B(-1)^{\ve_A},
\e{305}
where in turn $X^A$ now are classical fields. Notice that
$P_A=-i\hbar\rho^{-1/2}\d_A\circ\rho^{1/2}(-1)^{\ve_A}$ since the
eigenstates $|X\hb$
are normalized according to
\be
&&\int |X\hb\rho(X)dX \vb X|=\bett.
\e{306}
 The most general  $\Delta$-operator considered so far is $\Delta_0$
\cite{BTU}. We do not know to what extent a nonzero $G(X)$ may be used.
Notice that
the nilpotence of $\Delta$ requires $\Delta_0$ to be nilpotent and $G$ to
satisfy
\be
&&\Delta_0G=0, \quad \Delta_0\half(-1)^{\ve_A}\rho^{-1}\left(\d_A\rho
E^{AB}\right)-G\stackrel{\lea}{\d}_AE^{AB}=0.
\e{307}
The equivalence between \r{301} and \r{302} follows if we choose
\be
&&\vb X|\cW\hb=\exp{\{{i\over\hbar}\cW(X)\}} \quad \Lra \quad
|\cW\hb=\exp{\{{i\over\hbar}\cW(X)\}}\rho^{1/2}|0\hb_P.
\e{308}
Notice the normalization $\vb X|\rho^{1/2}|0\hb_P=1$.

The partition function $Z$, \ie the path
integral of the gauge fixed action, is  given by
$Z=\vb \cX|\cW\hb$, where $|\cW\hb$ is the master state and $|\cX\hb$ a
gauge fixing
state both satisfying the quantum master equation \r{302}.
$|\cW\hb$  and $|\cX\hb$ have the general form
\be
&&|\cW\hb=\exp{\left\{{i\over\hbar}\cW(X)\right\}}\rho^{1/2}|0\hb_{P\pi}, \quad
|\cX\hb=\exp{\left\{-{i\over\hbar}\cX\dagg(X,\la)\right\}}\rho^{1/2}|0\hb_{P
\pi},
\e{309}
 where $\la^\al$ are
Lagrange multipliers and
$\pi_\al$ their conjugate momenta. The vacuum state
$|0\hb_{P\pi}=|0\hb_P\otimes|0\hb_{\pi}$ satisfies
$P_A|0\hb_{P\pi}=\pi_\al|0\hb_{P\pi}=0$. By means of the extended eigenstates
$|X,\la\hb$ satisfying the completeness relations
\be
&&\int |X,\la\hb\rho(X)dX d\la\vb X,\la|=\bett,
\e{310}
and $\vb X,\la|\rho^{1/2}| 0\hb_{P,\la}=1$,
the partition function $Z=\vb \cX|\cW\hb$ becomes
explicitly
\be
&&Z=\vb \cX|\cW\hb=\int\rho(X)dX d\la
\exp{\left\{{i\over\hbar}\biggl[\cW(X)+\cX(X,\la)\biggr]\right\}}
\e{311}
where  $\cW$ and $\cX$ in the path integral denotes
the master action and gauge fixing actions,  $\vb
X,\la|\cW(X)\rho^{1/2}| 0\hb_{P,\la}$
 and $\vb X,\la|\cX(X,\la)\rho^{1/2}| 0\hb_{P,\la}$
respectively, which both
satisfy the quantum
master equation \r{301}. This agrees with the results in \cite{BTU,Trip}.
The path
integral \r{311} is invariant under the anticanonical transformation
\be
&&\del X^A=(X^A, -\cW+\cX)\mu,
\e{3111}
where $\mu$ is an odd constant. It is also invariant under changes of the gauge
fixing function $\cX$ in accordance with the general invariance of the
quantum master
equation. Infinitesimally we have invariance under
\be
&&\del\cX=(\cX, f)-i\hbar\Delta f,
\e{3112}
where $f$ is an odd function.

In order to illustrate the above
 generalized BV-quantization  we give a detailed    explicit treatment in
terms of the Darboux coordinates $x^a$ and $x^*_a$ \cite{Quanti}. In the
BV-quantization they are viewed as fields and antifields.   The nilpotent and
hermitian $Q$-operator in conventional BV-quantization is given by \r{107}.
Notice the relation
\be
&&\vb x, x^*|Q|S\hb=-(i\hbar)^2\Delta\vb x, x^*|S\hb,
\e{312}
where
\be
&&\Delta=(-1)^{\ve_a}{\d\over\d x^a}{\d\over\d x^*_a}
\e{313}
is  the well-known
nilpotent operator in the BV-quantization. The quantum master equation is
\be
&&Q|S\hb=0\quad \Lra \quad \Delta \exp\left\{{i\over\hbar}S(x,x^*)\right\}=0,
\e{314}
if
\be
&&|S\hb\equiv\exp{\left\{{i\over\hbar}S(x,x^*)\right\}}|0\hb_{pp^*}.
\e{315}

The path  integral of the gauge fixed action, $Z$, is  given by
$Z=\vb \Psi|S\hb$, where $|S\hb$ is the master state \r{315} and $|\Psi\hb$
a gauge
fixing state  also satisfying the quantum master equation \r{314}. \ie
$Q|\Psi\hb=0$. In the standard case
we have explicitly
\be
&&|\Psi\hb=\exp\left\{\hbar^{-2}[Q,
\Psi(x)]\right\}|0\hb_{px^*},
\e{316}
where the operators have the hermiticity properties
\be
&&(x^a)^{\dag}=x^a,\;\; (x^*_a)^{\dag}=-x^*_a, \;\;
(p_a)^{\dag}=p_a(-1)^{\ve_a}, \;\;
(p^a_*)^{\dag}=p_*^a(-1)^{\ve_a},  \;\; \Psi^{\dag}=\Psi,
\e{317}
 and where the vacuum states satisfy
\be
&&p_a|0\hb_{pp_*}=p_*^a|0\hb_{pp_*}=0, \quad
p_a|0\hb_{px^*}=x^*_a|0\hb_{px^*}=0, \quad Q|0\hb_{pp_*}=Q|0\hb_{px^*}=0.\nn\\
\e{318}
Note that $S(x,x^*)$ and $\Psi(x)$ belong to the class of commuting
operators. Note
also that $|\Psi\hb$ satisfies
\be
&& \left(x^*_a-\d_a\Psi(x)\right)|\Psi\hb=0, \quad
\left(p_a+p_*^b\d_b\d_a\Psi(x)\right)|\Psi\hb=0,
\e{319}
which fixes $p_a$ and $x^*_a$. The explicit form of the gauge fixed
partition function
is then
\be
&&Z=\vb \Psi|S\hb=\,_{px^*}\vb
0|\exp{\{\hbar^{-2}[ Q,
\Psi(x)]\}}\exp{\{i\hbar^{-1}S(x,x^*)\}}|0\hb_{pp_*}=\nn\\
&&=\int Dx Dx^*
\exp{\{i\hbar^{-1}S(x,x^*)\}}\del(x^*_a-\d_a\Psi(x)),
\e{320}
where the last equality is obtained by inserting the completeness
relations\footnote{Note that for odd $n$ the states in \r{321}, \r{322} do
not have a
definite Grassmann parity \cite{RM}.}
\be
&&\int |x, x^*\hb D x D x^* \vb x, x^*|=\int  |x, p_*\hb D x D p_*\vb x,
p_*|=\bett,
\e{321}
and the properties
\be
&&_{px^*}\vb 0|x, p_*\hb=\vb x,
x^*|0\hb_{pp_*}=1, \quad \vb
p_*|x^*\hb=(2\pi\hbar)^{-n_-/2}\hbar^{n_+/2}\exp{\{-i\hbar^{-1}p^a_*x^*_a\}}
,\nn\\
\e{322}
where $n_+$ ($n_-$) is the number of bosons (fermions) among the $x^a$
operators.
Eq.\r{320} agrees with the standard BV quantization \cite{BV}. The
independence of
the gauge fixing operator $\Psi$ follows from
\be
&&\del Z=\vb \Psi|\int_0^1d\al\,
\exp{\{-\al\hbar^{-2}[Q, \Psi]\}}\hbar^{-2}[Q,
\del\Psi]\exp{\{\al\hbar^{-2}[Q,
\Psi]\}}|S\hb=0,
\e{323}
since $|S\hb$ and $|\Psi\hb$ satisfy the master equation \r{314}.

\setcounter{equation}{0}
\section{Quantum antibrackets within general BFV-BRST quantization}
Within the Hamiltonian formulation of general gauge theories there are
first class
constraints $\theta_a=0$ where $\theta_a$ by  definition are variables in
arbitrary involution with  respect to the Poisson bracket on the considered
symplectic manifold, \ie
 \be
 &&\{\theta_a, \theta_b\}=U_{ab}^{\;\;\;c}\theta_c,
\e{401}
where the structure coefficients $U_{ab}^{\;\;\;c}$ may be arbitrary  functions.
In \cite{BFV} it was shown that the algebra \r{401} on a ghost extended manifold
always may be embedded in one single real, odd function $\Om$, the BFV-BRST
charge,
in such a way that $\{\Om, \Om\}=0$ in terms of the
 extended Poisson bracket.
The corresponding quantum theory
is consistent if the corresponding odd, hermitian operator
$\Om$ is nilpotent, \ie $\Om^2=0$. For a finite number of degrees of
freedom such a
solution always exists and is of the form \cite{BF} ($N$ is the rank of the
theory
and $\theta_a$ are the hermitian constraint operators)
\be
 &&\Om=\sum_{i=0}^N\Om_i,
 \e{402}
\be
&&\Om_0\equiv\ca^a\theta_a,\;\;\;\Om_i\equiv \Om_{a_1\cdots a_{i+1}}^{b_i\cdots
b_1}(\pet_{b_1}\cdots\pet_{b_i}\ca^{a_{i+1}}\cdots\ca^{a_1})_{Weyl},
\;\;\;i=1,\ldots,N,
\e{403}
where we have introduced the ghost operators $\cC^a$ and their conjugate momenta
$\pet_a$  satisfying
 \be
&&[\ca^a,
\pet_b]=i\hbar\del^a_b,\quad(\cC^a)\dagg=\cC^a,
\quad\pet_a\dagg=-(-1)^{\varepsilon_a}\pet_a.
\e{404}
The Grassmann parities are
\be
&&\ve_a\equiv \ve(\theta_a), \quad \ve(\cC^a)=\ve(\pet_a)=\ve_a+1.
\e{4041}
The operators $\Om_{a_1\cdots a_{i+1}}^{b_i\cdots
b_1}$ in \r{403}, whose explicit form we do not give here, contain the original
operators and are such that
$\Om$ is hermitian and nilpotent.
(In \r{403} the ghost operators are Weyl ordered which means that $\Om_i$
are all
hermitian.)
If the ghosts $\ca^a$ are assigned ghost number one and $\pet_a$ ghost
number minus
one,
$\Om$ in
\r{402} has ghost number one. Notice the relation
\be
&&[G, \Om]=i\hbar\Om,\quad
G\equiv-\half\left(\pet_a\cC^a-\cC^a\pet_a(-1)^{\ve_a}\right),
\e{405}
where $G$ is the ghost charge.

The BRST charge \r{402}-\r{403} is the BRST charge in
the minimal sector. In order to construct gauge fixed theories in the most
general
way we need to extend the manifold further with Lagrange multipliers and
antighosts. The complete BRST charge $Q$, which also is hermitian and nilpotent,
contains
$\Om$ and additional terms involving the Lagrange multipliers and antighosts
\cite{BF}. The gauge fixing is then performed by means of an odd operator
$\Psi$ such
that the effective gauge fixed Hamiltonian contains the operator $[Q,
\Psi]$. $\Psi$
seems always possible to be chosen to be nilpotent in such a way that it may be
identified with a coBRST charge
\cite{GMS}.

\subsection{Equations of motion in terms of antibrackets}
The  equations of motion in a gauge fixed theory is determined by
an effective Hamiltonian operator of the form $H=\cH+(i\hbar)^{-1}[Q,
\Psi]$ where
 $\cH$ commutes with $Q$. Now any gauge theory may be expressed in terms of an
equivalent reparametrization invariant theory in which the
Hamiltonian is of the form $H=(i\hbar)^{-1}[Q,
\Psi]$ where the new $Q$ contains a term with a new ghost multiplying the
constraint variable $P_0+\cH$, where $P_0$ is the conjugate momentum to
time, and
where  the new gauge fixing fermion
$\Psi$ has a corresponding new gauge fixing term.
The effective equations of motion becomes then
\be
&&\dot{A}=(i\hbar)^{-1}[A, H]=(i\hbar)^{-2}[A, [Q, \Psi]]=(i\hbar)^{-2} (A,
\Psi)_Q-(i\hbar)^{-2}\half[Q, [\Psi,A]],\nn\\
\e{406}
where the quantum antibracket is the general one defined by \r{1}.  The
last equality
implies
\be
&&\vb\phi|\biggl((i\hbar)^{2}\dot{A}-(A,
\Psi)_Q\biggr)|\phi'\hb=0,
\e{407}
where $|\phi\hb$, $|\phi'\hb$ are physical states annihilated by $Q$.
The  equation of motion \r{406} may also be expressed entirely in terms of
antibrackets if we make use of the symmetric combination of both $Q$- and
$\Psi$-antibrackets. We have the relation
\be
&&(i\hbar)^{2}\dot{A}=[A, [Q, \Psi]]={2\over3}\biggl( (A,
\Psi)_Q+(A, Q)_{\Psi}\biggr).
\e{4071}

\subsection{Quantum master equation for generalized Maurer-Cartan equations.}
In \cite{OG} we proposed a quantum master equation for generalized Maurer-Cartan
equations of operators in arbitrary involutions. Such operators are encoded
in the
hermitian, nilpotent BFV-BRST charge operator, which in the minimal sector
is given by
$\Om$ in \r{402}-\r{403}. Finite group transformations on operators are
determined by
the  Lie equations
\be
&&A(\phi)\stackrel{\lea}{\nabla}_a\equiv
A(\phi)\stackrel{\lea}{\d_a}-(i\hbar)^{-1}[A(\phi), Y_a(\phi)]=0,
\e{408}
where $\d_a$ is a derivative with respect to the parameter $\phi^a$,
$\ve(\phi^a)=\ve_a$. ($\phi^a$ may also be viewed as a new set of commuting
operators.) On states the corresponding Lie equations are
\be
&&\vb
A(\phi)|\stackrel{\lea}{D}_a\equiv\vb
A(\phi)|\left(\stackrel{\lea}{\d_a}-(i\hbar)^{-1} Y_a(\phi)\right)=0.
\e{4081}
The operator
$Y_a$ in both these relations depends on
$\phi^a$ and must satisfy the integrability conditions
\be
Y_a\stackrel{\lea}{\d_b}-Y_b\stackrel{\lea}{\d_a}(-1)^{\ve_a\ve_b}=
(i\hbar)^{-1}[Y_a,
Y_b],
\e{409}
which in turn are integrable without further conditions. In order for the
Lie equations \r{408} to be connected to the integration of the quantum
involution
encoded in $\Om$, $Y_a(\phi)$ has to be of the form
\be
&&Y_a(\phi)=\la^b_a(\phi)\theta_b(-1)^{\ve_a+\ve_b}+\{\mbox{\small possible ghost
dependent terms}\},\quad
\la^b_a(0)=\del^b_a,
\e{410}
where $\la^b_a(\phi)$ are operators in general. One may note that in a ghost
independent scheme
the generators of the finite transformations  are $\theta_a$. However,
within our
BRST framework, $[\Om, \pet_a]=\theta_a+\{${\small ghost dependent
terms}$\}$, are
the appropriate generators which motivates the form \r{410}.  Here we take
one step
further and define
$Y_a$ to be of the general form
\be
&&Y_a(\phi)=(i\hbar)^{-1}[\Om, \Om_a(\phi)],\quad\ve(\Om_a)=\ve_a+1,
\e{412}
where $Y_a$ has ghost number zero and $\Om_a$ ghost number minus one.
This form implies that $[Y_a(\phi), \Om]=0$, so that if $[A(0), \Om]=0$
then $[A(\phi), \Om]=0$, \ie a BRST invariant operator remains BRST
invariant when
 transformed according to
\r{408}.
Equations
\r{410} and
\r{412}  together with
\r{402} and \r{403}  imply that  $\Om_a$ in \r{412} must be of the form
\be
&&\Om_a(\phi)=\la^b_a(\phi)\pet_b+\{\mbox{\small possible ghost
dependent terms}\},\quad
\la^b_a(0)=\del^b_a.
\e{411}
One may note that the Lie equations \r{408} also may be defined in terms
antibrackets.
From \r{406} we have
\be
&&A(\phi)\stackrel{\lea}{\d_a}-(i\hbar)^{-2} (A,
\Om_{\al})_{\Om}-(i\hbar)^{-2}\half[[A, \Om_{\al}], \Om](-1)^{\ve_{\al}}=0,
\e{4121}
where we have introduced the general quantum antibracket defined in
accordance with
\r{1}.
The integrability condition \r{409} for $Y_a$ leads  by means of \r{412} to the
following equivalent equation for
$\Om_a$
\be
&&\Om_a\stackrel{\lea}{\d_b}-\Om_b\stackrel{\lea}{\d_a}(-1)^{\ve_a\ve_b}=
(i\hbar)^{-2}(\Om_a, \Om_b)_{\Om}-\half(i\hbar)^{-1}[\Om_{ab}, \Om],
\e{413}
which also involves the $\Om$-antibracket in \r{4121}.
 Due to the form
 \r{411} of $\Om_a$, eq.\r{413} are generalized Maurer-Cartan equations for
$\la^b_a(\phi)$. The integrability conditions of \r{413} lead to
equivalent first order equations for $\Om_{ab}$ and so on. Thus,  $Y_a$ is
replaced by a whole set of operators, and  the integrability conditions
\r{409} for
$Y_a$ are replaced by a whole set of integrability conditions.

In \cite{OG} we proposed that all these integrability conditions
are
embedded in one single  quantum master equation  given by
\be
&&(S, S)_{\Delta}=i\hbar[\Delta, S],
\e{418}
where $\Delta$ is  an extended nilpotent BFV-BRST charge given by
\be
&&\Delta\equiv\Om+\eta^a \pi_a(-1)^{\ve_a},\quad \Delta^2=0,\quad
[\phi^a, \pi_b]=i\hbar\del^a_b,
\e{415}
where in turn $\pi_a$
are conjugate momenta
 to
$\phi^a$,  now  turned into operators, and
$\eta^a$, $\ve(\eta^a)=\ve_a+1$, are new ghost variables to  be treated as
parameters.
The operator $S(\phi, \eta)$ in the master equation \r{418} is   an even
operator
defined by
\be
&&S(\phi,
\eta)\equiv G+\eta^a\Om_a(\phi)+\half\eta^b\eta^a\Om_{ab}(\phi)(-1)^{\ve_b}+
\nn\\&&+{1\over6}\eta^c\eta^b\eta^a\Om_{abc}(\phi)(-1)^{\ve_b+\ve_a\ve_c}+
\ldots\nn\\&&\ldots+
{1\over n!}\eta^{a_n}\cdots\eta^{a_1}\Om_{a_1\cdots
a_n}(\phi)(-1)^{\ve_n}+\ldots,\nn\\&& \ve_n\equiv \sum_{k=1}^{[{n\over
2}]}\ve_{a_{2k}}+\sum_{k=1}^{[{n-1\over
2}]}\ve_{a_{2k-1}}\ve_{a_{2k+1}},
\e{416}
where $G$ is the ghost charge operator in \r{405}. (In \cite{OG} we made another
choice for $\ve_n$.) The
operators
$\Om_{a_1\cdots a_n}(\phi)$ in \r{416} satisfy the properties
\be
&&\ve(\Om_{a_1\cdots a_n})=\ve_{a_1}+\ldots+\ve_{a_n}+n,\quad[G, \Om_{a_1\cdots
a_n}]=-n i\hbar\Om_{a_1\cdots a_n}.
\e{417}
The last relation implies that $\Om_{a_1\cdots a_n}$ has ghost number minus
$n$.
If we assign ghost
number one to $\eta^a$ then $\Delta$ has ghost number one and $S$ has ghost
number
zero.
Our main conjecture is that the
operators
$\Om_{a_1\cdots a_n}(\phi)$ in \r{416} may  be identified with $\Om_{a}$,
$\Om_{ab}$ in
\r{413} and all the $\Om$'s in their integrability conditions in a
particular manner.

The antibracket $(S, S)_{\Delta}$ in \r{418} is the quantum antibracket
defined by \r{1} with $Q$ replaced by $\Delta$. Thus, we have
\be
&&\quad (S, S)_{\Delta}\equiv[[ S, \Delta], S].
\e{4181}
By means of this relation it is easily seen that the consistency of \r{418} requires
$[\Delta, S]$ to be nilpotent. We have
\be
&&[\Delta, (S, S)_{\Delta}]=0\quad \Lra\quad [\Delta, S]^2=0.
\e{419}
The explicit form of $[S, \Delta]$ is to the lowest orders in $\eta^a$
\be
&&[S, \Delta]=i\hbar\Om+\eta^a[\Om_a,
\Om]+\eta^b\eta^a\Om_a\stackrel{\lea}{\d_b}i\hbar(-1)^{\ve_b}+
\half\eta^b\eta^a[\Om_{ab},
\Om](-1)^{\ve_b}+\nn\\&&+\half\eta^c\eta^b\eta^a\Om_{ab}
\stackrel{\lea}{\d_c}i\hbar(-1)^{\ve_b+\ve_c}+{1\over
6}\eta^c\eta^b\eta^a[\Om_{abc},
\Om](-1)^{\ve_b+\ve_a\ve_c}+O(\eta^4).
\e{4191}
To zeroth and first order in $\eta^a$ the master equation \r{418} is satisfied
identically. However, to second order in $\eta^a$ it yields exactly \r{413}.
 At  third order in $\eta^a$ it yields
\be
&&\d_a\Om_{bc}(-1)^{\ve_a\ve_c}+\half(i\hbar)^{-2}(\Om_a,
\Om_{bc})_{\Om}(-1)^{\ve_a\ve_c}+cycle(a,b,c)=\nn\\&&=-(i\hbar)^{-3}(\Om_a,
\Om_b,
\Om_c)_{\Om}(-1)^{\ve_a\ve_c}-{2\over3}(i\hbar)^{-1}[\Om'_{abc},\Om],\nn\\
&&
\Om'_{abc}\equiv\Om_{abc}-{1\over8}\left\{(i\hbar)^{-1}[\Om_{ab},
\Om_c](-1)^{\ve_a\ve_c}+cycle(a,b,c)\right\},
\e{420}
where
we have introduced the higher quantum antibracket of order 3 defined by  \r{206}
and explicitly given by \r{208} with
$Q$ replaced by $\Om$.

Comparing equation \r{420} and the integrability conditions of \r{413} we
find exact
agreement. We have also checked that the consistency condition
\r{419}  yields exactly
\r{409} to second order in $\eta^a$, which is consistent with \r{413} as it
should.
Similarly we have checked that \r{419} to third
order in $\eta^a$  yields a condition which is consistent with \r{420},
exactly like
\r{409} is consistent with \r{413}.

 The
master equation \r{418} yields at higher orders in $\eta^a$ equations
 involving still higher quantum
$\Om$-antibrackets defined by \r{206} and
operators
$\Om_{abc\ldots}$ with still more indices. We conjecture that these equations
agree exactly  with the integrability conditions of
\r{420}. For a rank-$N$ theory we expect that there exists a solution of
the form
\r{416} to the master equation \r{418}, which terminates just at the maximal
order $\eta^{N}$. In the appendix we treat quasigroup first rank theories
in detail.

We  end this subsection with some transformation formulas of the master equation
\r{418} \cite{Sp2QA}. Let us define the transformed operators
$S(\al)$ and $\Delta(\al)$ by
\be
&&S(\al)\equiv e^{{i\over\hbar}\al F}
Se^{-{i\over\hbar}\al F}, \quad
\Delta(\al)\equiv e^{{i\over\hbar}\al F}
{\Delta} e^{-{i\over\hbar}\al F},
\e{421}
where $\al$ is a parameter and $F$  an arbitrary even operator. If $S$ and
$\Delta$
satisfy the master equation \r{418} then  $S(\al)$ and $\Delta(\al)$ satisfy the
transformed master equation
\be
&&(S(\al), S(\al))_{\Delta(\al)}=
i\hbar[\Delta(\al), S(\al)].
\e{422}
If  $F$ in \r{421} is restricted to satisfy
the master equation \r{418}, \ie
\be
&&(F, F)_{\Delta}=i\hbar[\Delta, F],
\e{423}
then $\Delta(\al)$ in \r{421} reduces to
\be
&&\Delta(\al)=\Delta+(i\hbar)^{-1}
[\Delta, F](1-e^{-\al}).
\e{424}
(Notice that \r{423} implies $\Delta''(\al)+\Delta'(\al)=0$.) For $F=S$ we
have in
particular that $S$ satisfies the master equation \r{418}
with $\Delta$ replaced by $\Delta(\al)$ in \r{424} where $F$ is replaced by $S$.

There are also transformations on $S$
leaving $\Delta$ unaffected for which the
master equation \r{418} is invariant.
 The natural
automorphism of \r{418} is
\be
&&S\;\ra\;S'\equiv \exp{\biggl\{-(i\hbar)^{-2}[\Delta,
\Psi]\biggr\}}S\exp{\biggl\{(i\hbar)^{-2}[\Delta,
\Psi]\biggr\}},
\e{425}
where $\Psi$ is an arbitrary odd operator.
It is easily seen that $S'$ also satisfies
the master equation \r{418}. For infinitesimal transformations we have
\be
&&\del S=(i\hbar)^{-2}[S, [\Delta, \Psi]],\nn\\
&&\del_{21} S\equiv(\del_2\del_1-
\del_1\del_2)S=(i\hbar)^{-2}[S, [\Delta,
\Psi_{21}]],\nn\\
&&\Psi_{21}=(i\hbar)^{-2}(\Psi_2, \Psi_1)_{\Delta}.
\e{426}
Analogously to the equivalent forms of the general equations of motion in
\r{406} and
\r{4071} we have also
\be
&&\del S=(i\hbar)^{-2}\biggl( (S, \Psi)_{\Delta}-\half[\Delta, [\Psi,
S]]\biggr)=\nn\\&&=(i\hbar)^{-2}{2\over3}\biggl((S, \Psi)_{\Delta}+(S,
\Delta)_{\Psi}\biggr).
\e{427}

If the transformation \r{425} connects any solutions of the master equation
\r{418}
then the general solution is
\be
&&S=\exp{\biggl\{-(i\hbar)^{-2}[\Delta,
\Psi]\biggr\}}G\exp{\biggl\{(i\hbar)^{-2}[\Delta,
\Psi]\biggr\}},
\e{428}
where $\Psi$ depends on all variables including $\phi^a$ and $\eta^a$ but not on
$\pi_a$.
$\Psi$ is only required to have total ghost number minus one since $S$ has total
ghost number zero. The explicit form
\r{416} of $S$ is then reproduced.  Notice  that $S=G$ is a trivial
solution of the
master equation \r{418}.

We  would also like to mention that there is a possibility to extend the
formalism to
$\eta$-dependent states and operators which satisfy the Lie equations
\r{408}-\r{4081}
with the $\eta$-dependent connections $\tilde{Y}_a(\phi, \eta)\equiv
(i\hbar)^{-1}[\Delta,
\tilde{\Om}_a(\phi, \eta)]$ where $\tilde{\Om}_a(\phi, \eta)$ is determined
by the
equation
\be
&&S\stackrel{\lea}{\d_a}-(i\hbar)^{-1}[S, \tilde{Y}_a]=0.
\e{429}
These $\eta$-extended states, $|\tilde{A}\hb$, and operators, $\tilde{A}$,
satisfy
the equations
\be
&&S|\tilde{A}\hb=i\hbar gh(|A\hb)|\tilde{A}\hb, \quad [S,\tilde{A}]=i\hbar
gh(A)\tilde{A},
\e{430}
where $|A\hb=|\left.\tilde{A}\hb\right|_{\eta=0}$ and
$A=\left.\tilde{A}\right|_{\eta=0}$.   The details will be given
elsewhere.

\setcounter{equation}{0}
\section{Generalization to the Sp(2)-case}
There is an Sp(2)-extended version of BV-quantization \cite{Sp2,Sp2s} which
in its
most general form is called triplectic quantization \cite{Trip}. In this
formalism
there are two generalized antibrackets which are called Sp(2)-antibrackets.
Also these
brackets may be mapped on operators. The quantum Sp(2)-antibrackets are
defined by
\cite{Quanti,Sp2QA}
($a, b, c, \ldots=1,2$, are $Sp(2)$-indices which are raised (and lowered)
by the
Sp(2) metric $\ve^{ab}$ ($\ve_{ab}$))
\be
&&(f, g)^a_Q\equiv\half \left([f, [Q^a, g]]-[g, [Q^a,
f]](-1)^{(\ve_f+1)(\ve_g+1)}\right),
\e{501}
where $Q^a$ are two odd operators.
The corresponding classical antibrackets satisfy the properties \r{3}-\r{8}
except
that the Jacobi identities are  valid for symmetrized Sp(2)-indices. The quantum
Sp(2)-antibrackets \r{501} satisfy \r{3}-\r{6}. However, instead of the Jacobi
identities we have
\be
&&(f,(g,
h)^{\{a}_Q)^{b\}}_Q(-1)^{(\ve_f+1)(\ve_h+1)}+cycle(f,g,h)=\nn\\&&={1\over
6}(-1)^{\ve_f+\ve_g+\ve_h}\left\{\left([f, [g, [h, [Q^a, Q^b]]]]+\half[[f, [g,
[h, Q^{\{a}]]], Q^{b\}}]\right)(-1)^{\ve_f\ve_h}+\right.\nn\\
&&+\left.\left([f, [h,
[g, [Q^a, Q^b]]]]+\half[[f, [h,
[g, Q^{\{a}]]],
Q^{b\}}]\right)(-1)^{\ve_h(\ve_f+\ve_g)}\right\}+cycle(f,g,h),\nn\\
\e{502}
and instead of Leibniz' rule we have
\be
&&(fg, h)^a_Q-f(g, h)^a_Q-(f, h)^a_Qg(-1)^{\ve_g(\ve_h+1)}=\nn\\
&&=\half\left([f, h][g,
Q^a](-1)^{\ve_h(\ve_g+1)}+[f,Q^a][g,h](-1)^{\ve_g}\right).
\e{503}
The relation \r{13} generalizes \eg to
\be
&&(f, Q^{\{a})^{b\}}={3\over2}[f, [Q^a, Q^b]].
\e{504}
Higher order quantum Sp(2)-antibrackets are defined by \r{205} with $Q$
replaced by
$Q^a$. The properties \r{206}-\r{210} are then valid by the same
replacement. The
relations \r{2101}-\r{2102} are valid with two Sp(2)-indices. To obtain the
generalized Jacobi identities we need the corresponding identities to \r{211}.
However, since they involve the trivial equalities
\be
&&[e^{-A}Q^ae^A, e^{-A}Q^be^A]-e^{-A}[Q^a, Q^b]e^A=0,
\e{505}
which are symmetric in the Sp(2)-indices, the generalized Jacobi identities
corresponding to \r{2132} will also be symmetric in the Sp(2)-indices.
Corresponding to the treatment in section 4 we may also define ordinary quantum
Sp(2)-antibrackets by a restriction to a maximal set of commuting
operators. If these
operators are functions of commuting coordinate operators then $Q^a$ and
$[Q^a, Q^b]$
must be maximally quadratic in the momenta in order for the
Sp(2)-antibrackets to
strictly satisfy the Jacobi identities and Leibniz' rule.

In the considered applications of the quantum Sp(2)-antibrackets \r{501} to  the
Sp(2)-version of the BV-quantization and to  the
Sp(2)-version of BFV-BRST quantization the two odd operators
$Q^a$ were required to satisfy \cite{Sp2QA}
\be
&&Q^{\{a}Q^{b\}}\equiv Q^{a}Q^{b}+Q^{b}Q^{a}\equiv [Q^{a}, Q^{b}]=0.
\e{506}
The appearance of two  nilpotent defining operators is natural since the
Sp(2)-versions are directly related to the so called
BRST-antiBRST quantization
\cite{antiB}.
In the
Sp(2)-version of BV-quantization we have to consider commuting operators
which are
functions of a maximal commuting set of  coordinate operators.  The
operators $Q^a$
are then maximally quadratic in the momenta so that the quantum
Sp(2)-antibrackets
are ordinary ones. In the Schr\"odinger representation $Q^a$   are then
equal to the
two nilpotent differential operators, $\Delta^a$,  exactly like
$Q$ was represented by
$\Delta$ in section 5. The quantum master equations are here
\be
&&Q^a|\cW\hb=0, \quad a=1,2.
\e{507}
In
\cite{Sp2QA} this formalism was  shown to provide for an operator version of the
Sp(2)-extended BV-quantization corresponding to the what we had in section
5 for the
ordinary BV-quantization. For instance, the master equations and the gauge fixed
partition function in triplectic quantization were shown to follow from
\r{507} in
analogy to what we had in section 5.  The
quantum master equations  for generalized quantum Maurer-Cartan
equations for arbitrary open groups given in section 6 were also shown to be
possible to formulate in terms of the Sp(2)-brackets \r{501} in \cite{Sp2QA}.
\\ \\ \\
\noindent
{\bf Acknowledgments}

I.A.B. would like to thank Lars Brink for his very warm hospitality at the
Institute of Theoretical Physics, Chalmers and G\"oteborg University.   The
work of
I.A.B. is  supported by INTAS grant 96-0308
 and by RFBR grants 99-01-00980,
99-02-17916. I.A.B. and R.M. are thankful to the Royal Swedish Academy of
Sciences
for financial support.\\ \\ \\
\def\theequation{\thesection.\arabic{equation}}
\setcounter{section}{1}
\renewcommand{\thesection}{\Alph{section}}
\setcounter{equation}{0}
\noindent
{\Large{\bf{Appendix A}}}\\ \\
{\bf Application of the quantum master equation to quasigroup first rank
theories.}\\
\\ As an illustration of the formulas in subsection 6.2 we consider now
constraint
operators
$\theta_a$ forming a rank one theory in which case we have (we consider
$\ca\pet$-ordered operators here)
\be
&&\Om=\cC^a \theta'_a+\half \cC^b\cC^a
U^c_{ab}\pet_c(-1)^{\ve_c+\ve_b},\quad\theta'_a\equiv\theta_a+\half i\hbar
U_{ab}^b(-1)^{\ve_b}.
\e{a1}
The nilpotence of $\Om$ requires
\be
&&[\theta'_a, \theta'_b]=i\hbar U^c_{ab}\theta'_c,\nn\\
&&\left(i\hbar U^d_{ab}U^e_{dc}+[U^e_{ab},
\theta'_c](-1)^{\ve_c\ve_e}\right)(-1)^{\ve_a\ve_c}+{ cycle}(a,b,c)=0,\nn\\
&&(-1)^{\ve_a\ve_c}\left([U^e_{ab},
U^f_{cd}](-1)^{\ve_e(\ve_c+\ve_d)}-[U^f_{ab},
U^e_{cd}](-1)^{\ve_f(\ve_c+\ve_d)+\ve_e\ve_f}\right)+cycle(a,b,c)=0.\nn\\
\e{a2}
 The
last conditions are certainly satisfied if
\be
&&[U^c_{ab}, U^f_{de}]=0, \quad [[\theta_d,
U^c_{ab}], U^g_{ef}]=0,
\e{a3}
which corresponds to quasigroups \cite{Bat}.
 In this case $\Om_a$ may be chosen to be
\be
&&\Om_a(\phi)=\la^b_a(\phi)\pet_b,\quad \la^b_a(0)=\del^b_a,
\e{a4}
where we assume that
\be
&&[\la^b_a, \la^d_c]=0 \quad\Ra\quad[\Om_a, \Om_b]=0.
\e{a5}
The quantum $\Om$-antibracket  is then given by
\be
&&(\Om_a, \Om_b)_{\Om}=[\Om_a, [\Om, \Om_b]]=-(i\hbar)^2\la_a^f\la_b^e
U_{ef}^d\pet_d(-1)^{\ve_d+\ve_e+\ve_f+\ve_b\ve_f}+\nn\\&&+i\hbar
\left(\la^c_a[\theta'_c, \la^d_b ]-\la^c_b[\theta'_c,
\la^d_a](-1)^{\ve_a\ve_b}\right)\pet_d(-1)^{\ve_c}-\nn\\&&
-i\hbar\left(
\la^f_a\cC^e[U^c_{ef},
\la^d_b](-1)^{\ve_b(\ve_c+1)}-\la^f_b\cC^e[U^c_{ef},
\la^d_a](-1)^{\ve_a(\ve_c+1)+\ve_a\ve_b}\right)\pet_d\pet_c-\nn\\&&-
\cC^e[[\theta'_e, \la_b^c], \la^d_a
]\pet_d\pet_c(-1)^{(\ve_a+1)(\ve_b+\ve_c+1)}-
\nn\\&&-\half \cC^f\cC^e[[U^c_{ef},
\la^d_b], \la^g_a]\pet_g\pet_d\pet_c
(-1)^{\ve_c+\ve_f+(\ve_a+1)(\ve_b+\ve_c+\ve_d)+(\ve_b+1)(\ve_c+1)}.
\e{a6}
If we also require
\be
&&(\Om_a, \Om_b, \Om_c)_{\Om}=0\quad \Lra \quad [(\Om_a, \Om_b)_{\Om}, \Om_c]=0,
\e{a7}
then $(\Om_a, \Om_b)_{\Om}$ in \r{a6} satisfies the Jacobi identities which
makes
\r{413} integrable if $\Om_{ab}=0$. This condition is satisfied if we impose
\be
&&[\la^b_a,
U^c_{de}]=0,
\quad [\la^b_a,[\la^d_c,
\theta_e]]=0.
\e{a8}
Eq.\r{413} may now be written as
\be
&&\d_a\tilde{\la}_b^c-\d_b\tilde{\la}_a^c(-1)^{\ve_a\ve_b}=\tilde{\la}^e_a
\tilde{\la}^d_b
\tilde{U}^c_{de}(-1)^{\ve_b\ve_e+\ve_c+\ve_d+\ve_e}.
\e{a9}
where $\tilde{\la}_a^b\equiv V\la_a^bV^{-1}$ and $\tilde{U}_{ab}^c \equiv
VU_{ab}^cV^{-1}$ where in turn the operator $V(\phi)$ is determined by the
equation
\be
&&i \hbar \d_a V = V \lambda^b_a \theta'_b (-1)^{\ve_b}.
\e{a10}
 Eq.\r{a7} and
$\Om_{ab}=0$ make all higher integrability conditions identically zero.
One may note that
\be
&&Y_a(\phi)=(i\hbar)^{-1}[\Om, \Om_a]=\la^b_a
\theta'_b(-1)^{\ve_a+\ve_b}+\nn\\&&+
\la^b_a\cC^dU^c_{db}\pet_c(-1)^{\ve_a+\ve_c}+(i\hbar)^{-1}\cC^b[\theta_b,
\la^c_a]\pet_c.
\e{a11}

\end{document}